\documentclass[a4paper,11pt]{article}
\usepackage{pos}
\usepackage{hyperref}
\newcommand{\ben}{\begin{enumerate}}
\newcommand{\een}{\end{enumerate}}
\newcommand{\beq}{\begin{equation}}
\newcommand{\eeq}{\end{equation}}
\newcommand{\bal}{\begin{align}}
\newcommand{\eal}{\end{align}}

\newcommand{\bea}{\begin{eqnarray}}
\newcommand{\eea}{\end{eqnarray}}
\newcommand{\nn}{\nonumber\\}

\newcommand{\df}{{\rm d}}
\def\Dm1{{{\delta(1-z)}}}

\def\g0#1DY{{g_{0#1}^{DY}}}

\def\LogmW1{{{\ln (1-\omega)}}}

\newcommand{\overbar}[1]{\,\overline{\!{#1}}}
\newcommand{\Nbar}{\overbar{N}}
\newcommand{\gbar}{\overbar{g}}
\newcommand{\as}{a_s}
\newcommand{\muf}{\mu_F}

\newcommand{\eq}[1]{Eq.\ (\ref{#1})}
\newcommand{\fig}[1]{Fig.\ \ref{#1}}

\newcommand{\sect}[1]{Sec.\ \ref{#1}}

\title{Threshold Resummation of Drell-Yan type colorless processes at LHC}
\author[a,b]{Goutam Das}
\author*[c]{Chinmoy Dey}
\author[d]{M C Kumar}
\author[e]{Kajal Samanta}

\affiliation[a]{Institut f{\"u}r Theoretische Teilchenphysik und Kosmologie,\\ RWTH Aachen University, D-52056 Aachen, Germany}
\affiliation[b]{Department of Physics, Indian Institute of Technology Kanpur, Kanpur-208016, India}

\affiliation[c]{Theoretical Physics Division, Physical Research Laboratory,\\ Navrangpura, Ahmedabad 380009, India}

\affiliation[d]{Department of Physics, Indian Institute of Technology Guwahati,\\ Guwahati-781039, Assam, India}

\affiliation[e]{Institute for Particle Physics Phenomenology,\\ Durham University, Durham DH1 3LE, United Kingdom}

\emailAdd{gkdgoutam@iitk.ac.in}
\emailAdd{chinmoy@prl.res.in}
\emailAdd{mckumar@iitg.ac.in}
\emailAdd{kajal.samanta@durham.ac.uk}

\abstract{We look at the threshold effects in neutral and charged Drell-Yan production, Higgs boson production with a massive vector boson, and Higgs production in bottom quark annihilation at the Large Hadron Collider (LHC), up to the third order in QCD. Using third-order soft-virtual (SV) results and the universal properties of threshold logarithms, we find the process-dependent coefficients and improve the accuracy by including large threshold logarithms up to next-to-next-to-next-to-leading logarithmic (N$^3$LL) order and matched with the latest N$^3$LO results. 
We also show numerical results for the invariant mass distributions and total production cross sections for these processes. Our findings show that the theoretical scale uncertainties, which are about $0.4\%$ at N$^3$LO in fixed-order calculations, decrease to less than $0.1\%$ at N$^3$LO+N$^3$LL after SV threshold resummation in the high invariant mass region.}

\FullConference{17th International Symposium on Radiative Corrections: Applications of Quantum Field Theory to Phenomenology (RADCOR2025)\\
5-10 October 2025\\
Puri, India\\}

\begin{document}
{\flushright{TTK-26-03, IPPP/26/02}}
\maketitle

\section{Introduction}
Colorless production processes are key probes of particle properties, proton structure, and fundamental parameters at hadron colliders such as the LHC. Owing to their clean experimental signatures and relatively simple theoretical understanding, they play a central role in precision studies.
Prominent examples include Drell–Yan (DY) production, which serves as a standard candle for luminosity measurements, and Higgs-strahlung processes, where a Higgs boson is produced in association with a vector boson ($V=Z,W$), providing essential information on Higgs properties and possible physics beyond the Standard Model (BSM).
In this work, we consider neutral and charged DY production, associated $VH$ production, and Higgs production via bottom-quark annihilation, all of which are initiated by quark–antiquark annihilation at the Born level.

Fixed-order (FO) QCD corrections to DY production are known up to NNLO \cite{Hamberg:1990np} and have recently been extended to N$^3$LO \cite{Duhr:2020seh,Duhr:2020sdp}, with public codes now available \cite{Baglio:2022wzu}.
%
Near partonic threshold ($z\to1$), FO predictions suffer from large logarithmic enhancements that limit their reliability. Due to the universal structure of these terms, they can be resummed to all orders, thereby improving perturbative convergence and reducing scale uncertainties. The soft–virtual contributions required for threshold resummation are known through N$^3$LO and beyond \cite{Anastasiou:2014vaa,Catani:2014uta,Kumar:2014uwa,Das:2020adl,Ajjath:2020rci,Das:2022zie, Bhattacharya:2025rqk}, enabling resummation up to N$^3$LL and even N$^4$LL accuracy for selected processes.

To obtain accurate predictions across the full kinematic range, resummed results must be consistently matched with FO calculations. This is now possible using the N$^3$LO code \texttt{n3loxs} \cite{Baglio:2022wzu}.
The goal of this article is to present a systematic study combining N$^3$LL resummation with N$^3$LO fixed-order results for several color-singlet production processes at the LHC. The required resummation ingredients are extracted using the formalism given in Refs.~\cite{Ravindran:2005vv,Ravindran:2006cg}.

The paper is organized as follows: the theoretical framework is summarized in \sect{sec:theory}, numerical results are presented in \sect{sec:numerics}, and conclusions are given in \sect{sec:conclusion}.
\section{Theoretical framework}\label{sec:theory}
The hadronic cross-section for colorless
production
at hadron collider is given by,
\begin{align}\label{eq:had-xsect}
	\sigma(Q^2)
	&=\sum_{a,b={q,\overline q,g}} 
	\int_0^1 \df x_1
	\int_0^1 \df x_2~ f_a(x_1,\muf^2) ~
	f_b(x_2,\muf^2) 
	 \int_0^1 \df z \,\,
	\hat{\sigma}_{ab}(z,Q^2,\muf^2)
	\delta(\tau-z x_1 x_2)\,,
\end{align}
where $\sigma(Q^2) \equiv Q^2 \mathrm{d}\sigma/\mathrm{d}Q^2$ for DY-type processes, and $\sigma(Q^2) \equiv \sigma(M_H^2)$ in the case of the $b\bar{b}H$ process.
For the total production cross section of $VH$, this quantity is integrated over the invariant mass $Q$ of the final-state $VH$ system.
The hadronic and partonic threshold variables $\tau$ and $z$ are defined as
\begin{align}
\tau=\frac{Q^2}{S}, \qquad z= \frac{Q^2}{\hat{s}} \,,
\end{align}
where $S$ and $\hat{s}$ denote the hadronic and partonic center-of-mass energies, respectively.
The variables $\tau$ and $z$ are therefore connected through the relation $\tau = x_1 x_2 z$.
The partonic coefficient $\hat{\sigma}_{ab}$ can be further expressed in the following form,
\begin{align}\label{eq:partonic-decompose}
	\hat{\sigma}_{ab}(z,Q^2,\mu_F^2)
	=
	\sigma^{(0)}(Q^2) \Big[ 
		\Delta_{ab}^{\rm sv}\left(z,\muf^2\right) 
		+ \Delta_{ab}^{\rm reg}\left(z,\muf^2\right)
			\Big] \,.
\end{align}
The quantity $\Delta_{ab}^{\rm sv}$ is referred to as the soft–virtual (SV) partonic coefficient and encompasses all singular contributions arising in the $z \to 1$ limit. The term $\Delta_{ab}^{\rm reg}$, on the other hand, includes contributions that are regular in the variable $z$.
The overall normalization factor $\sigma^{(0)}$ is process dependent. For the specific processes considered here, it is given by the following expressions,
\begin{align}
	\sigma_{DY}^{(0)}(Q^2) 
	&= 
	\frac{\pi }{n_c } 
	\bigg[ {\cal F}_{DY}^{(0)}(Q^2) \bigg]\,, 
	\qquad \text{with } DY \in \{nDY, cDY, ZH, WH \} \,,
	\nn
	\sigma_{b\bar{b}H}^{(0)} 
	&= 
	\frac{\pi m_b^{2}(\mu_R^2) \tau}{6 M_H^2 v^2} \,,
\end{align}
where,
\begin{align}
	{\cal F}_{nDY}^{(0)}(Q^2) =&
	{4 \alpha^2 \over 3 S} \Bigg[Q_q^2 
	- {2 Q^2 (Q^2-M_Z^2) \over  \left((Q^2-M_Z^2)^2
	+ M_Z^2 \Gamma_Z^2\right) c_w^2 s_w^2} Q_q g_e^V g_q^V 
\nn
	&+ {Q^4 \over  \left((Q^2-M_Z^2)^2+M_Z^2 \Gamma_Z^2\right) c_w^4 s_w^4}\Big((g_e^V)^2
	+ (g_e^A)^2\Big)\Big((g_q^V)^2+(g_q^A)^2\Big) \Bigg]\,,
\nn
	{\cal F}_{cDY}^{(0)}(Q^2) =&
	{4 \alpha^2 \over 3 S} \Bigg[
	{Q^4 |V_{qq^{'}}|^2 \over  \left((Q^2-M_W^2)^2+M_W^2 \Gamma_W^2\right) s_w^4}\Big((g_e^{'V})^2
	+ (g_e^{'A})^2\Big)\Big((g_q^{'V})^2+(g_q^{'A})^2\Big) \Bigg]\,,
\nn
	\mathcal{F}^{(0)}_{ZH}(Q^2) =& 
	\frac{\alpha^2}{S}\Bigg[ 
		\frac{M_Z^2Q^2\lambda^{1/2}(Q^2,M_H^2,M_Z^2)
		\bigg( 1 + \frac{\lambda(Q^2,M_H^2,M_Z^2)}{12 M_Z^2/Q^2}\bigg)}{((Q^2-M_Z^2)^2+ M_Z^2\Gamma_Z^2)c_w^4s_w^4}
        \bigg((g_q^V)^2+(g_q^A)^2 \bigg)
	\Bigg]\,,
\nn
	\mathcal{F}^{(0)}_{WH}(Q^2) =& 
	\frac{\alpha^2}{S}\Bigg[ 
		\frac{M_W^2Q^2 |V_{qq^{'}}|^2\lambda^{1/2}(Q^2,M_H^2,M_W^2)
		\bigg( 1 + \frac{\lambda(Q^2,M_H^2,M_W^2)}{12 M_W^2/Q^2}\bigg)}{((Q^2-M_W^2)^2+M_W^2\Gamma_W^2)s_w^4}
        \bigg((g_q^{'V})^2+(g_q^{'A})^2 \bigg)
	\Bigg]\,.
\end{align}
Here, $V_{qq^{'}}$ denote the CKM matrix elements satisfying
$Q_{q} + Q_{q'}=\pm 1$, and
\begin{align}
g_a^A =& -\frac{1}{2} T_a^3 , \qquad
g_a^V = \frac{1}{2} T_a^3 - s_w^2 Q_a ,
\nn
g_a^{'A} =& -\frac{1}{2\sqrt{2}}, \qquad
g_a^{'V} = \frac{1}{2\sqrt{2}},
\end{align}
where $Q_a$ and $T_a^3$ correspond to the electric charge and weak isospin of the fermions, respectively. The parameters $M_V$ and $M_H$ represent the masses of the weak gauge boson and the Higgs boson, while $m_b$ denotes the bottom-quark mass and $v$ is the vacuum expectation value.
The function $\lambda$ appearing in the $VH$ channel is defined as
\begin{align}
\lambda(z,y,x) =
\bigg(1-\frac{x}{z}-
\frac{y}{z} \bigg)^2 -4\frac{xy}{z^2}.
\end{align}
The singular component of the partonic coefficient exhibits a universal behavior and receives contributions from the underlying hard form factor,
the mass factorization kernels
and soft radiation effects.
Each of these contributions is infrared divergent; however, after regularization and appropriate combination, they yield finite results. The finite singular terms possess a universal structure characterized by $\delta(1-z)$ and plus-distributions
${\cal D}_i = [\ln(1-z)^i/(1-z)]+$.
These logarithmically enhanced terms can be resummed to all perturbative orders in the threshold limit ($z \to 1$).
Threshold resummation is most conveniently carried out in Mellin ($N$) space, where convolution integrals reduce to simple products.

The partonic coefficient in Mellin space is organized as
\begin{align}\label{eq:resum-partonic}
\hat{\sigma}_N^{\rm N^{n}LL}
= \int_0^1 \df z ~ z^{N-1} \Delta^{\rm sv}(z)
\equiv g_{0} \exp \left( G_N \right) \,.
\end{align}
The factor $g_{0}$ is independent of the Mellin moment, while the resummation of threshold-enhanced logarithms is encoded in the exponent $G_{N}$. The logarithmic accuracy is determined by the successive terms in $G_N$, which up to N$^3$LL can be written as
\begin{align}\label{eq:gn}
	G_N = 
	\ln (\Nbar) ~\gbar_1(\Nbar)
	+\gbar_2(\Nbar)
	+\as ~\gbar_3(\Nbar)
	+ \as^2~\gbar_4(\Nbar)\,,
\end{align}
where $\Nbar = N \exp (\gamma_E)$ and $a_s$ is related to the strong coupling ($\alpha_s$) by $a_s = \alpha_s/(4\pi)$.
These coefficients are universal and depend only on whether the initial-state partons are quarks or gluons. Their explicit expressions can be found, for example, in \cite{Das:2022zie}.

To achieve full resummed accuracy, the $N$-independent coefficient $g_0$ must also be known to the appropriate perturbative order. In particular, up to N$^3$LL it is given by
\begin{align}\label{eq:g0}
	g_0
	=
	1
	+ \as ~g_{_{01}}
	+ \as^2 ~g_{_{02}}
	+ \as^3 ~g_{_{03}} \,.
\end{align}
The coefficients $g_{0i}$ are provided in Appendix of Ref. \cite{Das:2022zie}.

To express the resummed result in $z$ space, a Mellin inversion must be carried out, leading to
\begin{align}
\sigma^{\rm N^{n}LL}
=&
\sigma^{(0)} (Q^2)
\sum_{a,b \in {q,\bar{q}}}
\int_{c-i\infty}^{c+i\infty}
\frac{\df N}{2\pi i}
\tau^{-N}
f_{a,N}(\muf)
f_{b,N}(\muf)
~ \hat{\sigma}_N^{\rm N^{n}LL} .
\end{align}
During the evaluation of this complex integral, a Landau pole appears at
$N =\exp\big(1/(2 a_S \beta_0) -\gamma_E\big)$, which makes the choice of integration contour particularly important. We adopt the \textit{minimal prescription} as mention in Ref.~\cite{Catani:1996yz} by selecting the constant $c$ such that the Landau pole lies to the right of the contour, while all remaining singularities are located to its left.
The Mellin inversion is then performed along the contour defined by
$N = c + x~\exp(i\phi)$, with $x$ a real parameter and $c$ and $\phi$ specifying the contour. In this work, we choose $c=1.9$ and $\phi=3\pi/4$, which ensures numerical stability.
The matched cross-section can finally be written as
\begin{align}
\sigma^{\rm N^{n}LO+N^{n}LL}
=&
\sigma^{\rm N^{n}LO} 
+
\sigma^{(0)} (Q^2)
\sum_{a,b \in {q,\bar{q}}}
\int_{c-i\infty}^{c+i\infty}
\frac{\df N}{2\pi i}
\tau^{-N}
f_{a,N}(\muf)
f_{b,N}(\muf)
\bigg(
\hat{\sigma}_N^{\rm N^{n}LL}
-
\hat{\sigma}_N^{\rm N^{n}LL} \bigg|_{\rm tr}
\bigg) .
\end{align}
The quantities $f_{a,N}$ denote the Mellin moments of the parton distribution functions, defined in analogy with the partonic coefficient in \eq{eq:resum-partonic}. 
The subtraction term in the parentheses indicates that the resummed partonic coefficient in \eq{eq:resum-partonic} has been truncated to fixed order, thereby avoiding double counting of singular contributions already included in $\sigma^{\rm N^{n}LO}$.

\section{Numerical results}\label{sec:numerics}
In this section, we present numerical results for the color-singlet production processes discussed in the previous section, focusing on LHC phenomenology. Unless stated otherwise, all numerical results are obtained using the MMHT2014 \cite{Harland-Lang:2014zoa} PDFs via the {\tt LHAPDF} interface~\cite{Buckley:2014ana}. The LO and NLO cross sections are computed with the MMHT2014lo68cl and MMHT2014nlo68cl PDF sets, respectively, while the NNLO and N$^3$LO results use the MMHT2014nnlo68cl PDFs, with the central set (iset=0) adopted throughout.
The strong coupling $\as$ is chosen consistently with the {\tt n3loxs} code and varies order by order. The electromagnetic coupling is fixed to $\alpha \simeq 1/132.184142$. We take $m_Z = 91.1876$~GeV and $m_W = 80.379$~GeV, with widths $\Gamma_Z = 2.4952$~GeV and $\Gamma_W = 2.085$~GeV. The Weinberg angle $\sin^2\theta_{\rm w} = (1 - m_W^2/m_Z^2)$ is computed internally, corresponding to $G_F \simeq 1.166379\times 10^{-5} \text{GeV}^{-2}$.
The Higgs boson mass is set to $m_H = 125.1$~GeV with $v = 246.221$~GeV. The bottom- and top-quark pole masses are $m_b = 4.78$~GeV and $m_t = 172.76$~GeV, while their running masses are $m_b(m_b) = 4.18$~GeV and $m_t(m_t) = 162.7$~GeV. Unless specified otherwise, the proton–proton center-of-mass energy is $13$~TeV.

The central renormalization and factorization scales are chosen as $\mu_R = \mu_F = Q$, where $Q$ denotes the invariant mass of the final state. For Higgs production in bottom-quark annihilation, we instead use $\mu_R = m_H$ and $\mu_F = m_H/4$ following Ref.~\cite{Duhr:2019kwi}. Scale uncertainties are estimated by varying the unphysical scales subject to $|\ln(\mu_R/\mu_F)| \le \ln 2$, and are quoted as the maximum symmetric deviation from the central value.
To assess the impact of higher-order fixed-order and resummed corrections, we define the following ratios of cross sections, which are useful for experimental analyses:
\begin{eqnarray}
\text{K}_{\text{N}^i\text{LO}} = \frac{\sigma_{\text{N}^i\text{LO}}}{\sigma_{\text{LO}}} 
\quad \text{ and } \quad 
\text{R}_{\text{ij}} = \frac{\sigma_{\text{N}^i\text{LO} + \text{N}^i\text{LL}}}{\sigma_{\text{N}^j\text{LO}}} 
\quad \text{ with } 
\quad i, \, j= 0, 1, 2 \text{ and } 3 \, \cdot
	\label{eq:ratio}
\end{eqnarray}

In this section, we present the numerical results for Higgs production in association with a massive vector boson, $V = Z,\, W^-,\, W^+$. 
However, the invariant mass distributions are computed only for the Drell--Yan (DY) channel, where an off-shell gauge boson is produced first and then decays into an on-shell vector boson $V$ and a Higgs boson $H$.
In this work, we extend the invariant mass distribution 
calculation to third order, namely N$^3$LO+N$^3$LL accuracy. 
For this purpose, we use the N$^3$LO Drell--Yan invariant 
mass distributions provided by the {\tt n3loxs} code 
\cite{Baglio:2022wzu}. Instead of considering the decay of 
the off-shell vector boson into leptons, we implement 
numerically its decay into an on-shell vector boson $V$ 
and a Higgs boson $H$, following Eqs.~(2) and (3) of 
Ref.~\cite{Harlander:2018yio}. As a cross-check, we first 
reproduce the $VH$ invariant mass results from 
{\tt vh@nnlo 2.1} up to NNLO accuracy. After validating 
our setup, we extend the fixed-order predictions to 
N$^3$LO. For the resummed results, we employ our own 
numerical code, developed in a similar way to the one 
used earlier for dilepton production.

\begin{figure}[ht!]
	\centerline{
		\includegraphics[width=5.5cm, height=5.5cm]{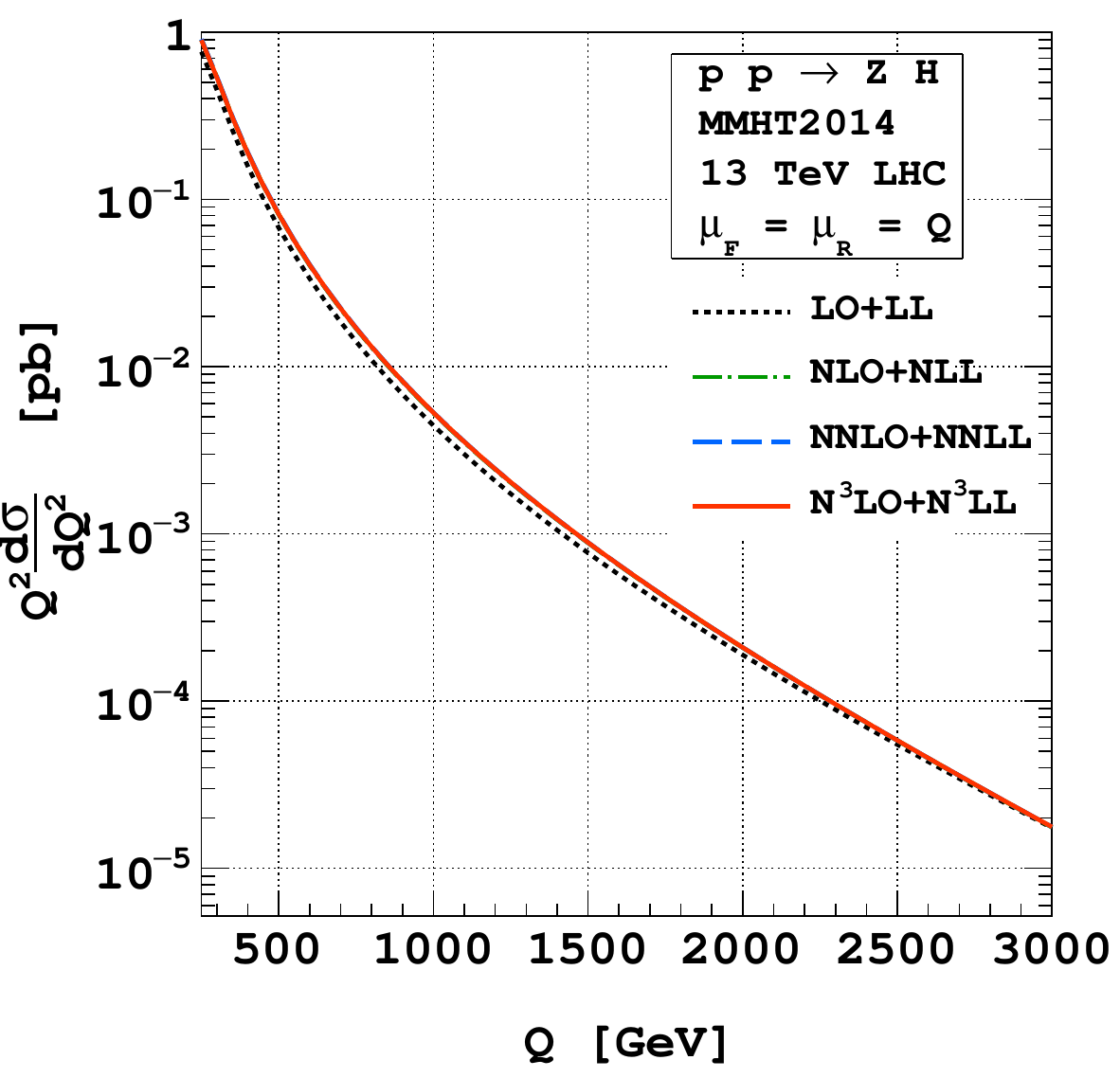}
             \includegraphics[width=5.5cm, height=5.5cm]{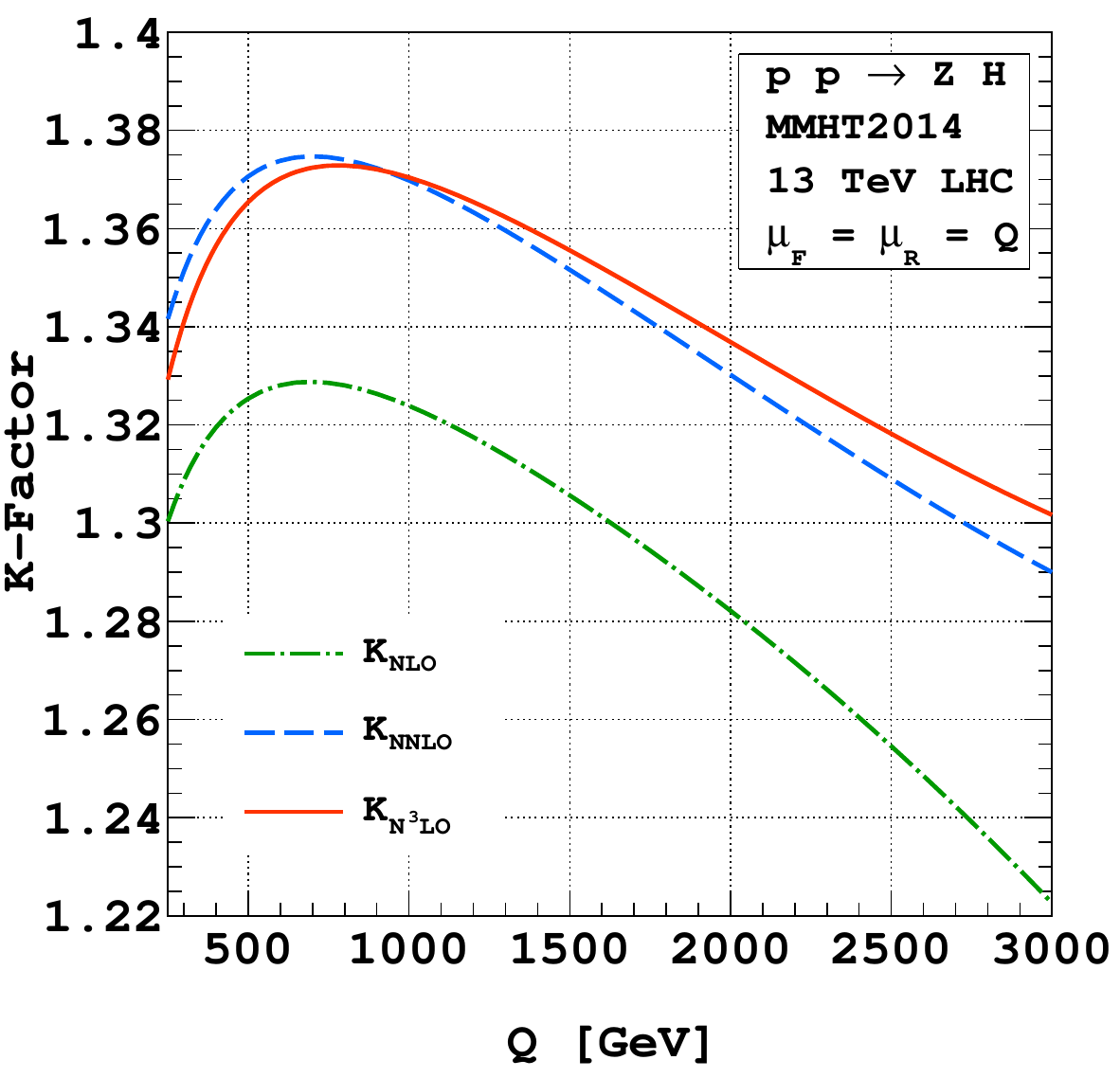}
             \includegraphics[width=5.5cm, height=5.5cm]{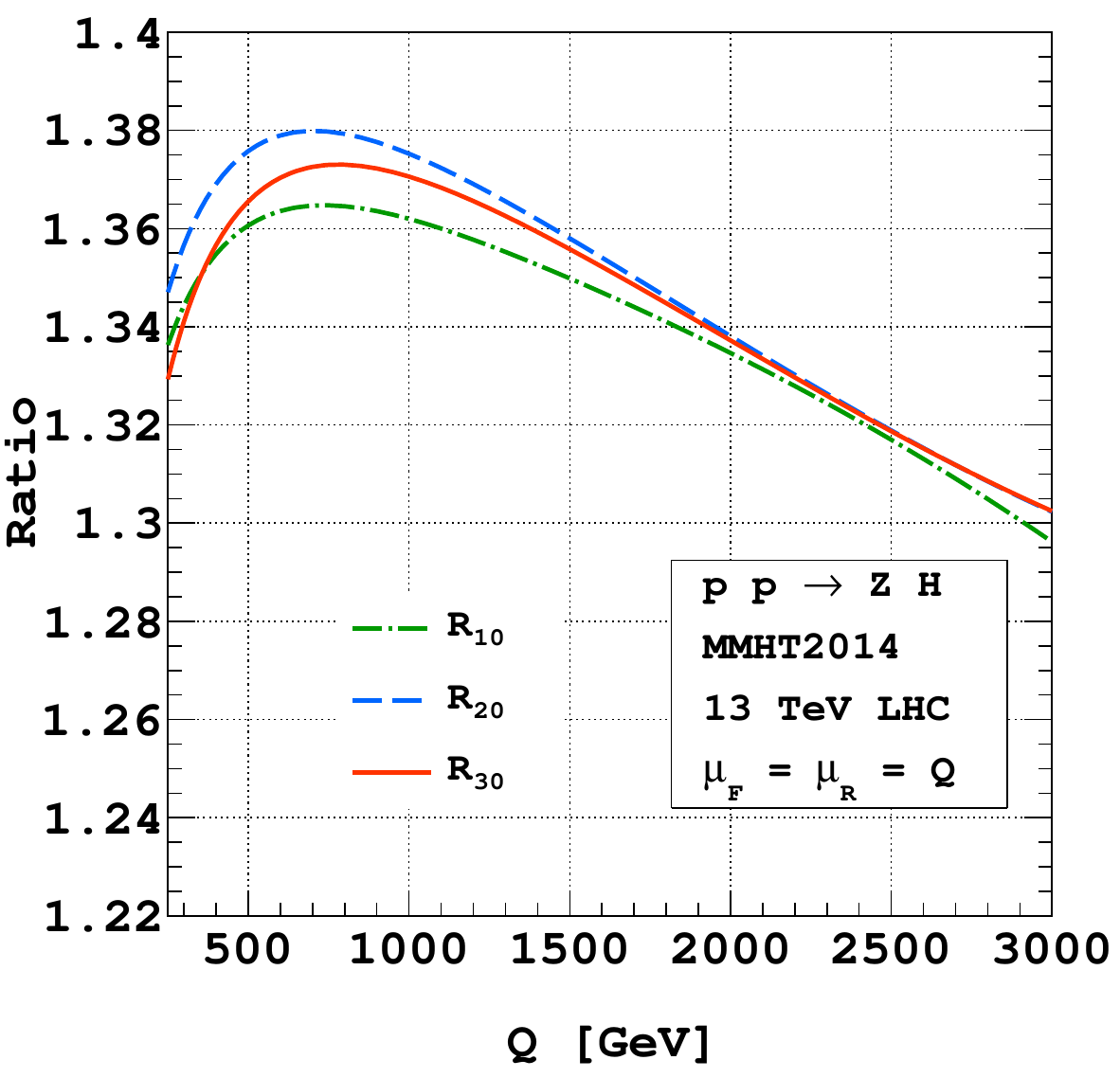}
	}
	\vspace{-2mm}
	\caption{\small{Invariant mass distribution of 
	$ZH$ for $13$ TeV LHC resummed (left) with
	fixed order (middle) and the resummed (right) K-factor.}}
	\label{fig:fo_resum_ZH}
\end{figure}

In~\fig{fig:fo_resum_ZH}, we present the invariant 
mass distribution of $VH$ production for the $ZH$ channel. 
The left panel shows the resummed results, while middle panel
shows the FO $K$-factor ratio up to N$^3$LO accuracy and the right panel displays the resummed $R$-factor ratio predictions up to N$^3$LO+N$^3$LL accuracy. The invariant mass $Q$ 
is varied between $250$ GeV and $3000$ GeV. The $K$-factors and 
$R$-factors defined in \eq{eq:ratio} are expected to be 
very similar to those of the neutral Drell--Yan case, 
because the branching contribution cancels in these ratios.
The corresponding $K_{N^iLO}$ and $R_{i0}$ factors, as defined earlier, 
are shown in the middle and right panel. We observe that the $K$-factor $K_{\rm NLO}$, 
$K_{\rm NLO}$ and $K_{\rm N^3LO}$ are 1.223, 1.290, and 1.302 respectively at $Q = 3000$ GeV. 
Whereas we observe that the $R$-factor $R_{10}$, 
$R_{20}$ and $R_{30}$ are 1.296, 1.302, and 1.302 respectively at $Q = 3000$ GeV.
These results indicate a very good perturbative convergence 
in the resummed predictions. In particular, the $R$-factors 
stabilize already at lower orders, whereas the fixed-order 
$K$-factors approach their asymptotic value more gradually. 
This behavior suggests that the resummed series exhibits 
faster convergence compared to the fixed-order expansion.

\begin{figure}[ht!]
	\centerline{
             \includegraphics[width=5.5cm, height=5.5cm]{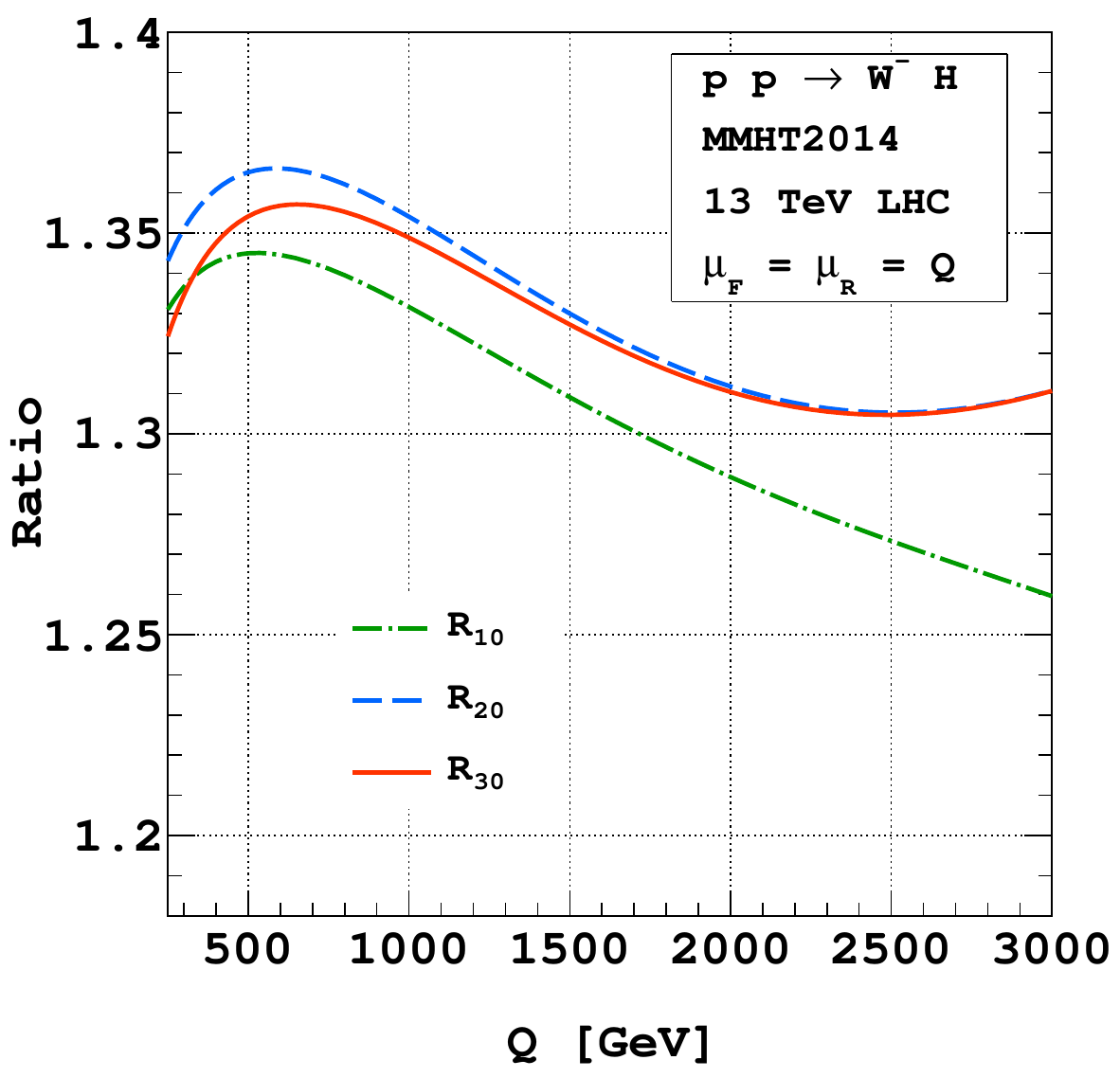}
             \includegraphics[width=5.5cm, height=5.5cm]{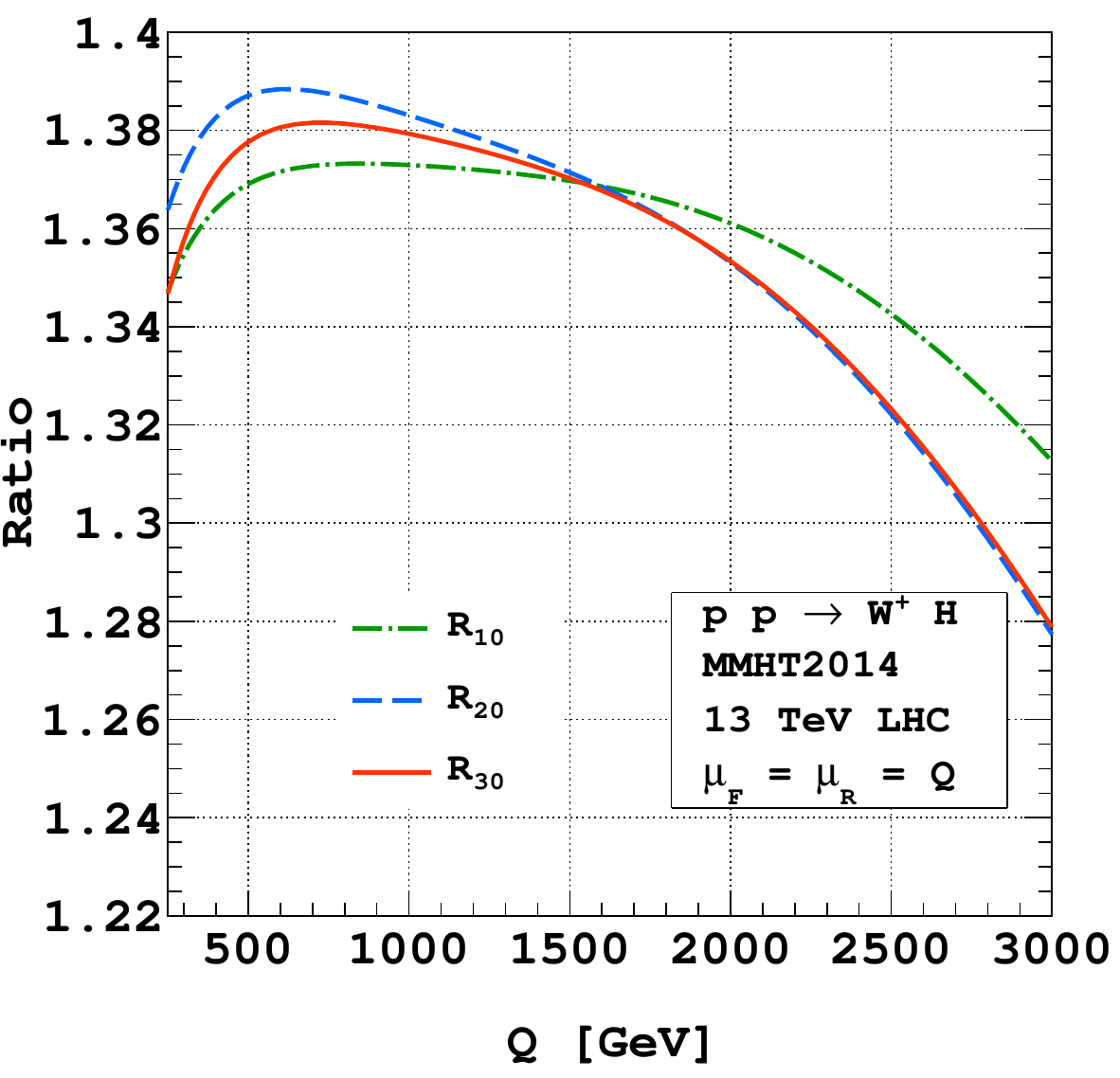}
	}
	\vspace{-2mm}
	\caption{\small{Resummed K-factor of 
	$WH$ for $13$ TeV LHC $W^-H$ (left) and $W^+H$ (right).}}
	\label{fig:fo_resum_WH}
\end{figure}

For completeness, ~\fig{fig:fo_resum_WH} presents 
the resummed $R$-factor ratios up to N$^3$LO+N$^3$LL accuracy for $WH$. In the left panel, we show the $R$-factor for $W^-H$ and the right panel displays the resummed $R$-factor ratio predictions up to N$^3$LO+N$^3$LL accuracy for $W^+H$.
We observe a behavior 
similar to that found in the $ZH$ case. The perturbative 
series shows good convergence in the resummed predictions. 
In particular, the $R$-factors reach a stable value already 
at lower perturbative orders, while the fixed-order 
$K$-factors approach their limiting value more slowly as 
the order increases. This pattern again indicates that the resummed expansion 
converges faster than the fixed-order series, demonstrating 
the improved perturbative stability of the resummed results.

\begin{figure}[ht!]
	\centerline{
		\includegraphics[width=5.5cm, height=5.5cm]{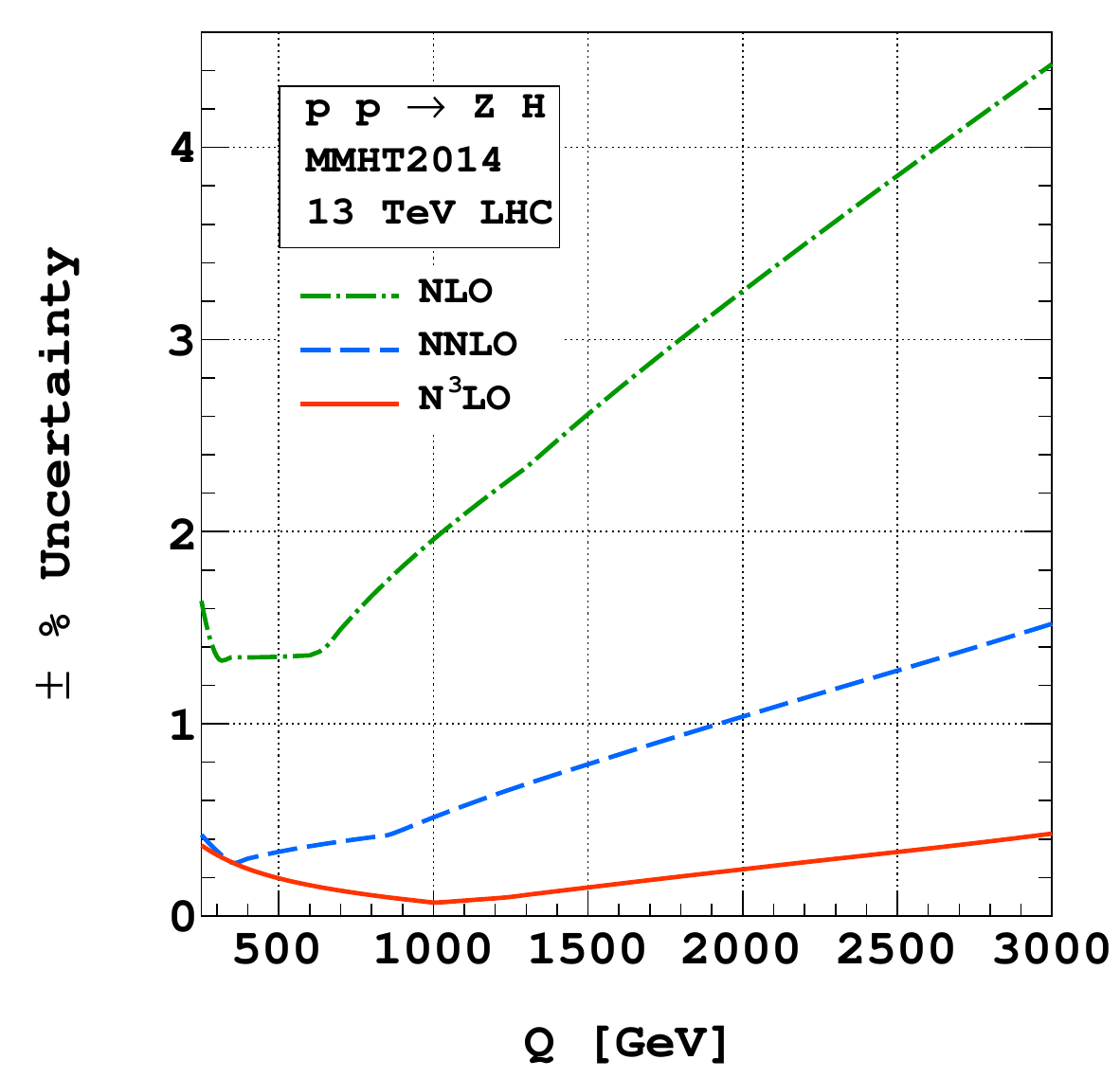}
		\includegraphics[width=5.5cm, height=5.5cm]{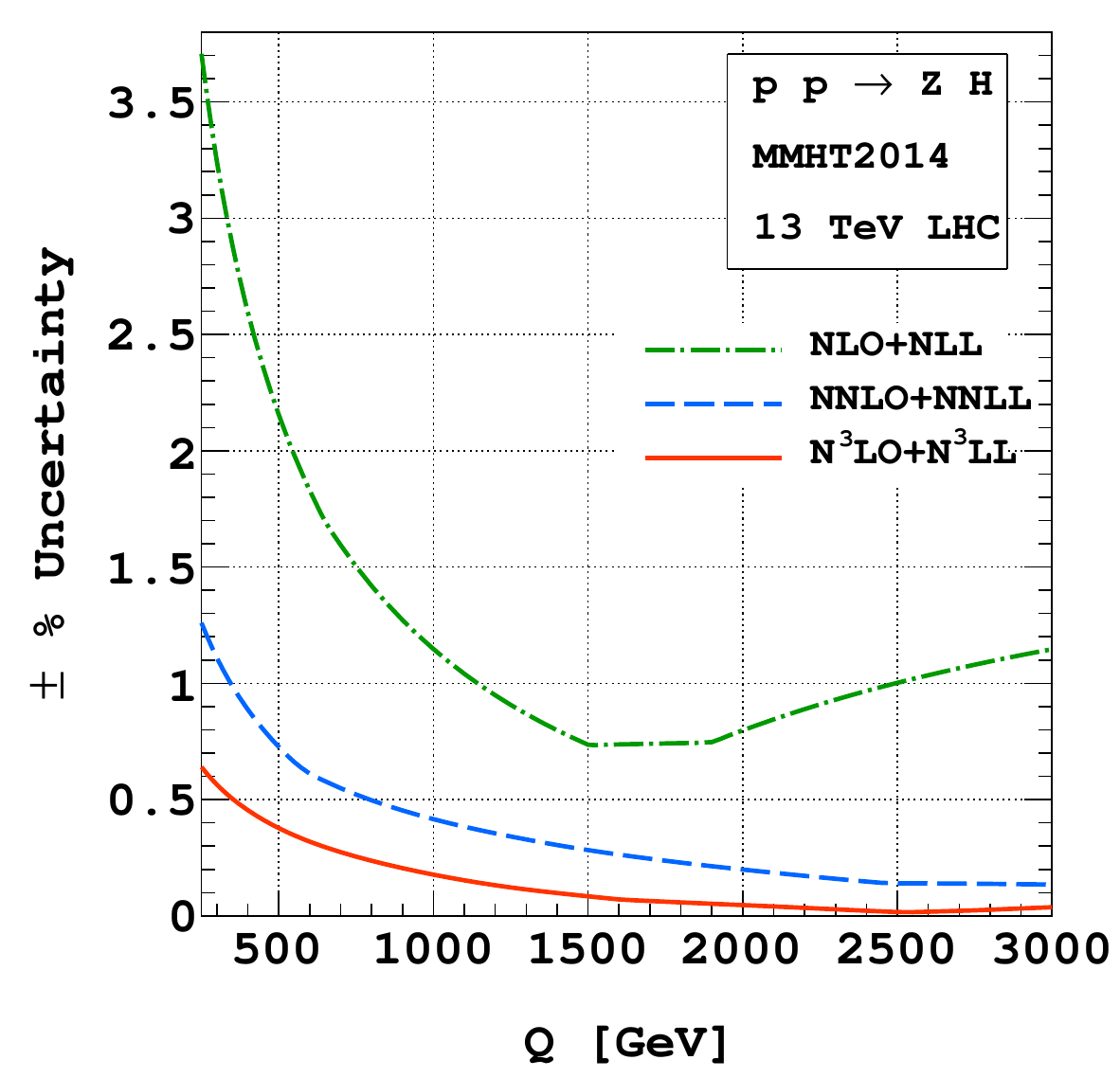}
		\includegraphics[width=5.5cm, height=5.5cm]{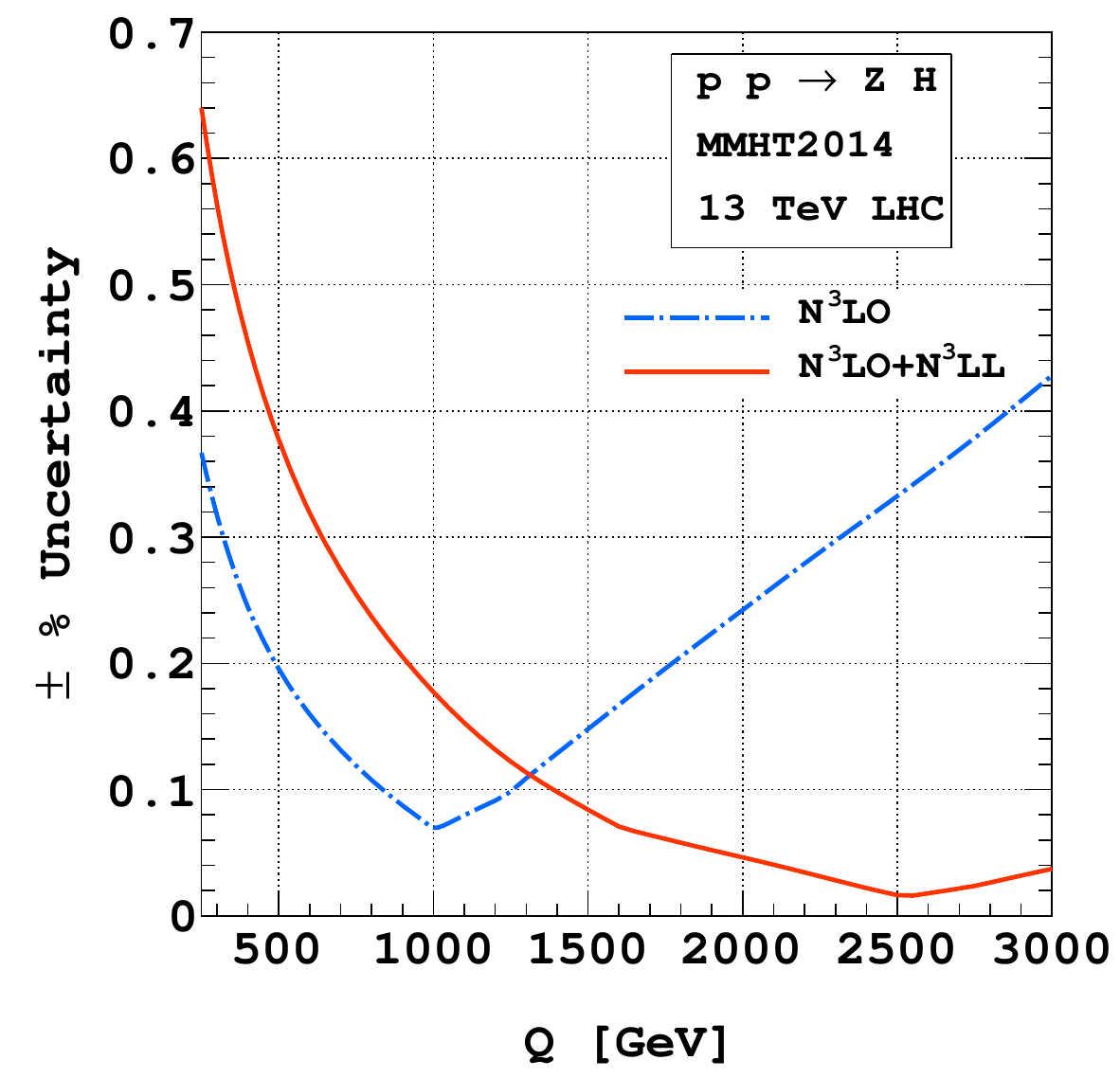}
	}
	\vspace{-2mm}
	\caption{\small{The $7$-point scale uncertainty for $ZH$, FO (left panel), resummed (middle panel) and comparison between FO and resummed (right panel).}}
	\label{fig:comparison_uncertainty}
\end{figure}

Besides, the convergence of the perturbation theory, the resummed results play a 
significant role in reducing the conventional $7$-point 
scale uncertainties. 
It is important to emphasize that resummation does not 
necessarily reduce the scale uncertainties in the low 
invariant mass region, for example below $1000$ GeV. 
In this region, threshold logarithms are not the only 
dominant contributions to the cross section, and other 
regular terms as well as different partonic channels 
play a significant role. As a result, the improvement 
over fixed-order (FO) predictions is not clearly visible 
there.
On the other hand, in the high $Q$ region, the resummed 
predictions show a noticeable reduction in scale 
uncertainties. To illustrate this behavior, we compare 
the standard 7-point scale variations for FO and 
resummed results at third order in QCD, as shown in 
~\fig{fig:comparison_uncertainty}. We find 
that at low $Q$, where non-logarithmic terms and 
additional partonic contributions are relevant, the FO 
results have smaller uncertainties. In contrast, at 
large $Q$, where threshold logarithms become dominant, 
the resummed results lead to smaller scale uncertainty and reduced to less than 0.1\%.
We also find that for fixed-order results, the scale 
uncertainties increase slowly as the invariant mass $Q$ 
becomes large. On the other hand, in the resummed results, 
the scale uncertainties decrease with increasing $Q$. This 
happens because, at large $Q$, the cross section is mainly 
controlled by threshold logarithms, which are resummed to 
all orders in perturbation theory.

\begin{figure}[ht!]
        \centerline{
                \includegraphics[width=5.5cm, height=5.5cm]{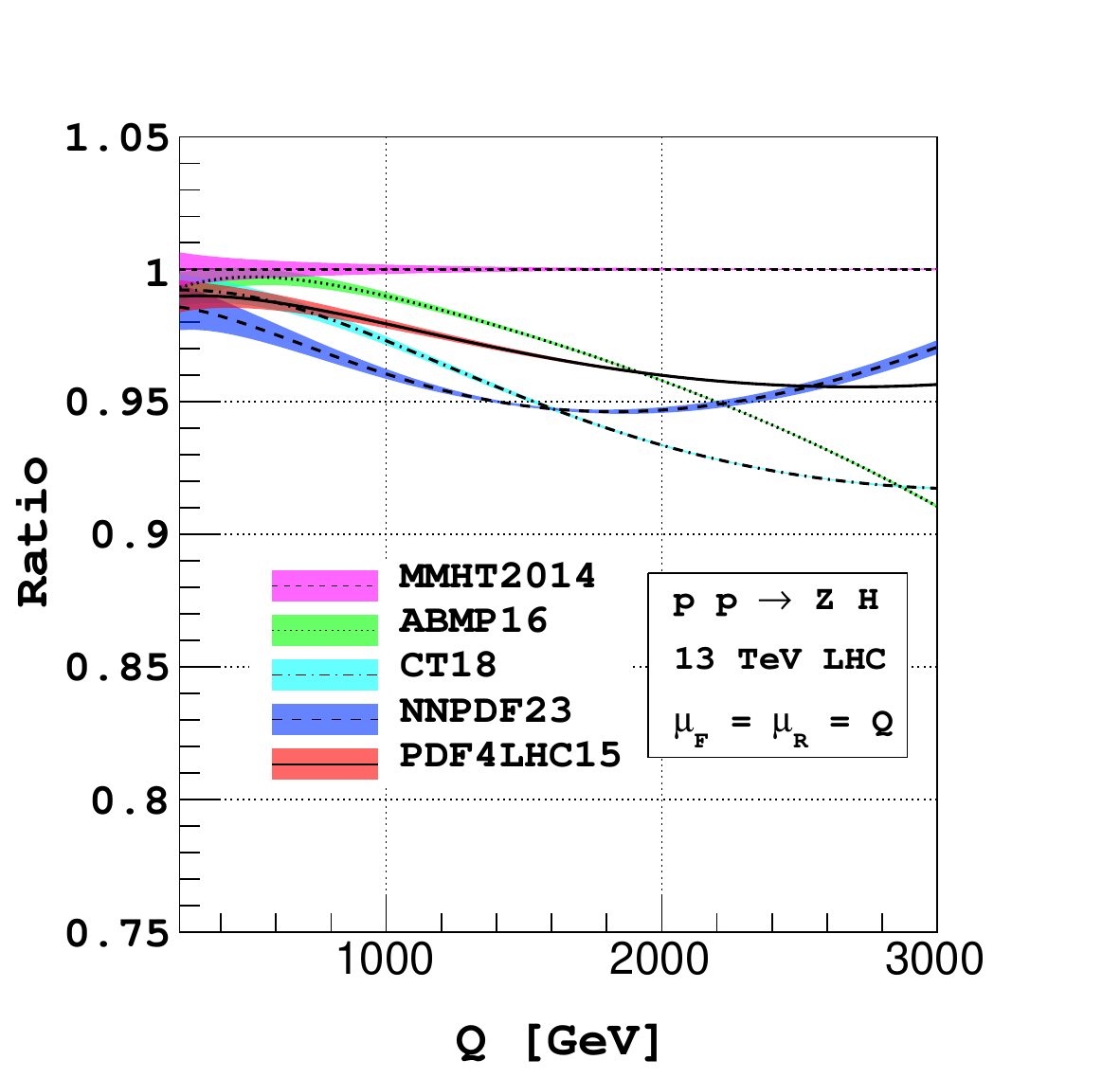}
                \includegraphics[width=5.5cm, height=5.5cm]{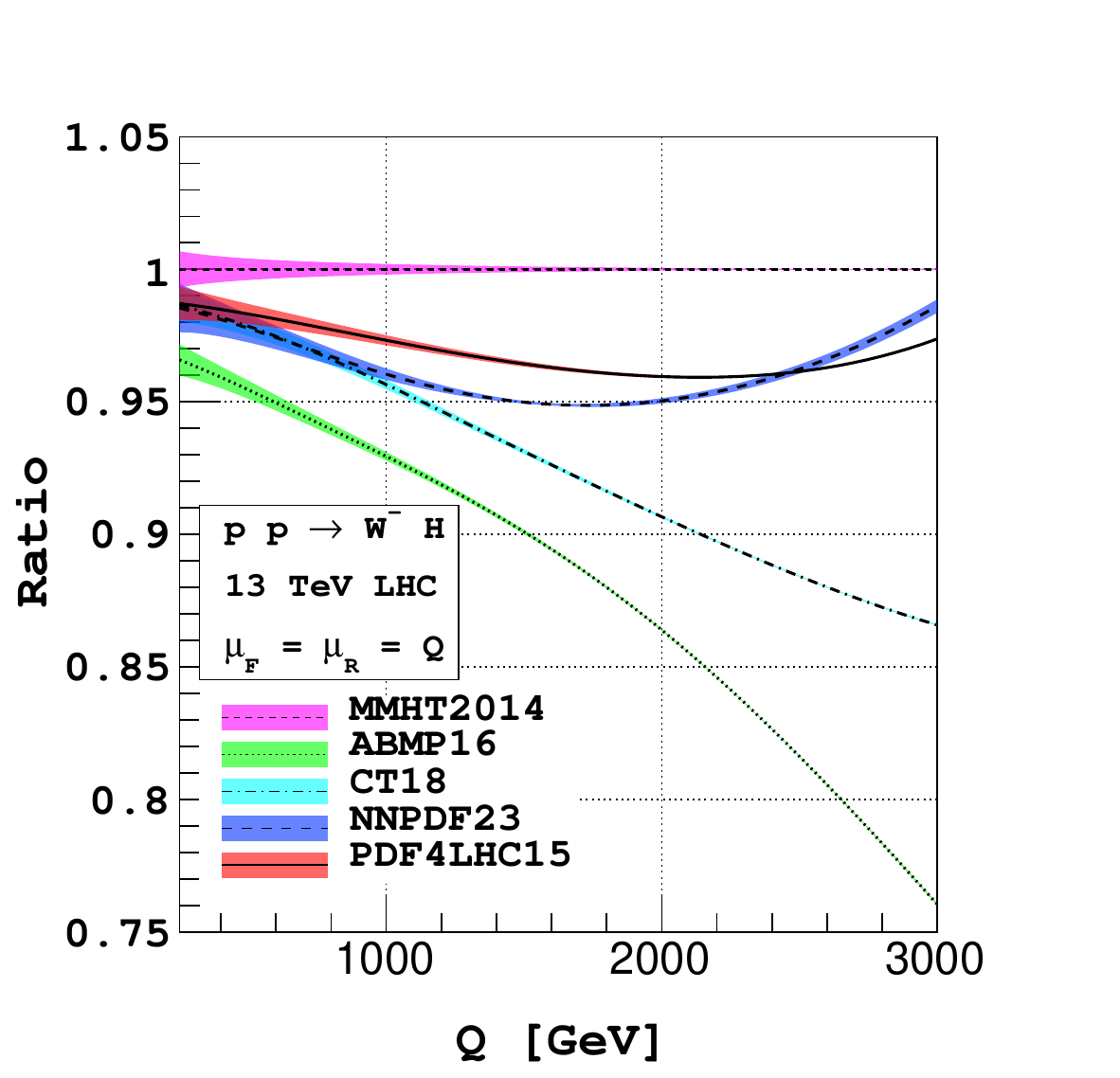}
                \includegraphics[width=5.5cm, height=5.5cm]{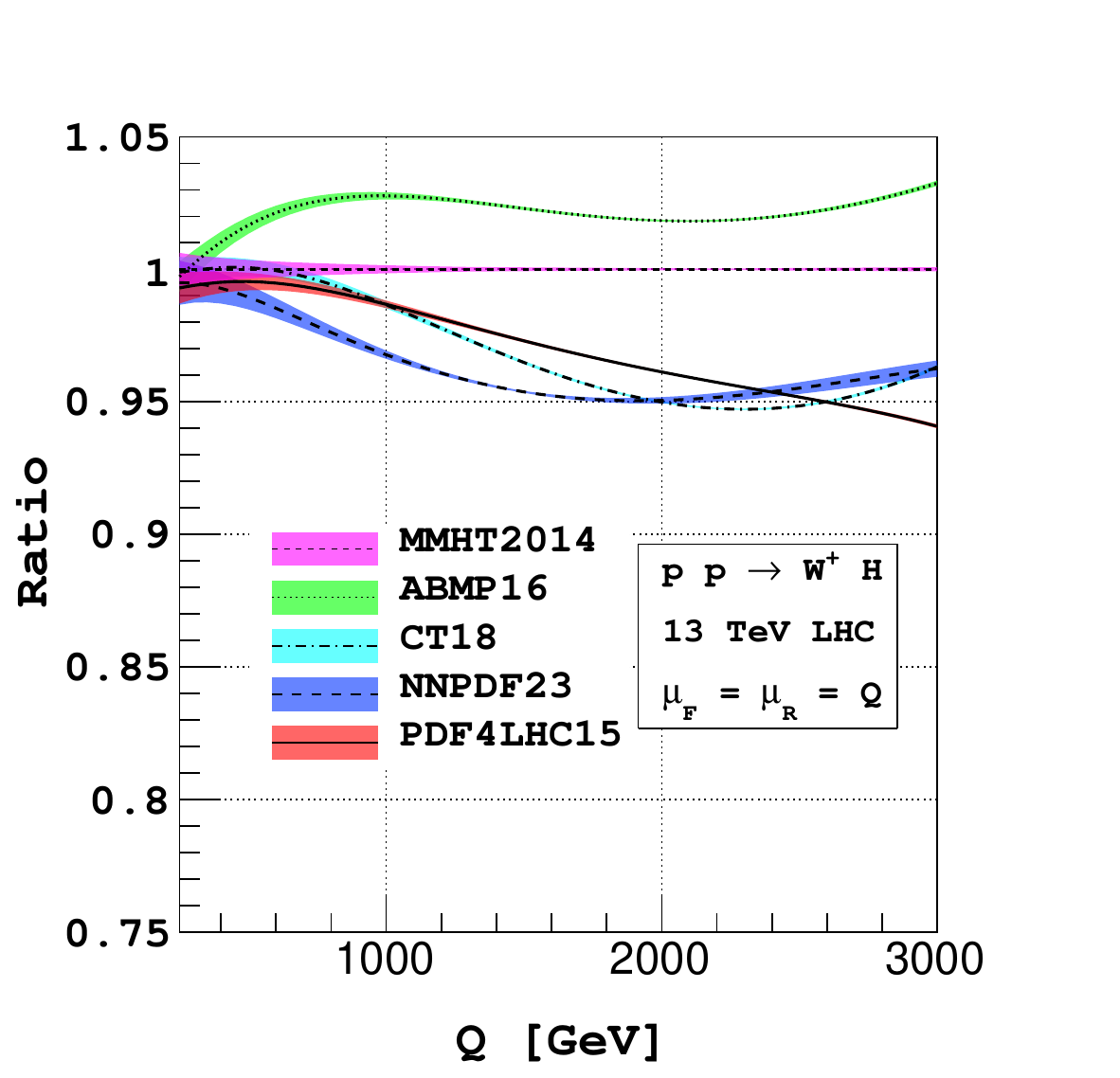}
        }
        \vspace{-2mm}
	\caption{
		\small{
			Behavior of invariant mass 
			distributions for different PDF groups at N$^{3}$LO+N$^{3}$LL 
			for $ZH$ (left panel), W$^-$H (middle panel) 
			and W$^+$H (right panel) with 7-point scale variation in band and
			(central value) normalized to the default 
			choice of MMHT2014. }
			}
        \label{fig:pdf_uncertainty_comp}
\end{figure}
\begin{figure}[ht!]
	\centerline{
		\includegraphics[width=7.0cm, height=7.0cm]{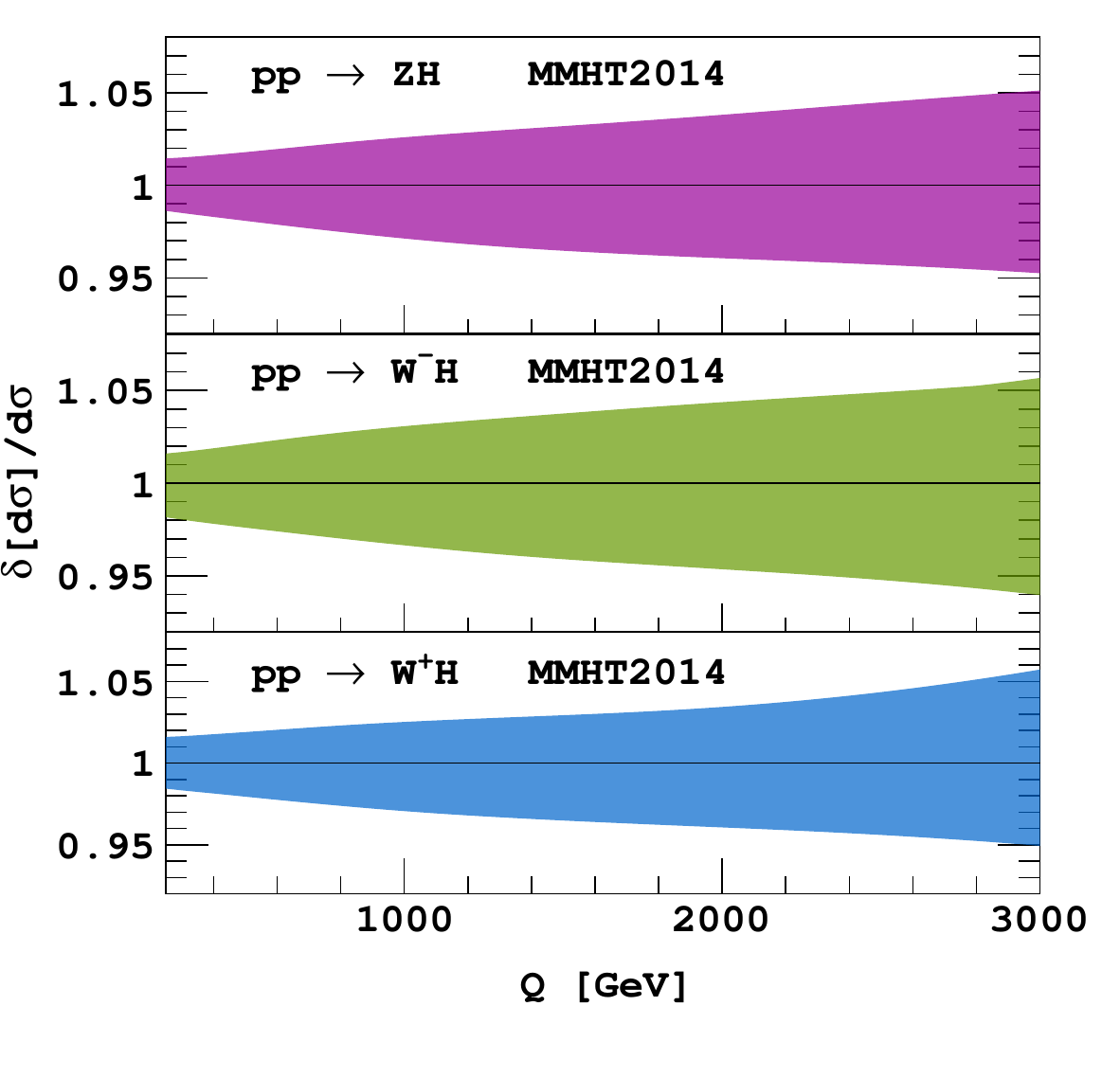}
	}
	\vspace{-2mm}
	\caption{\small{The intrinsic PDF 
	uncertainties in the 
	resummed results for $VH$ 
	production processes 
at N$^3$LO+N$^3$LL, obtained from 
MMHT2014nnlo68cl PDF sets.}}
	\label{fig:match_PDFuncertainty}
\end{figure}

Finally, we examine the sensitivity of the $VH$ invariant 
mass distributions to the choice of parton distribution 
functions (PDFs). The N$^3$LO+N$^3$LL predictions are 
computed using the central members (iset=0) of the NNLO 
PDF sets ABMP16~\cite{Alekhin:2017kpj}, CT18~\cite{Hou:2019qau}, 
NNPDF23~\cite{Ball:2012cx}, and PDF4LHC15~\cite{Butterworth:2015oua}. 
All results are normalized to our default MMHT2014nnlo 
central set and evaluated at the central scale. The 
comparison of different PDF uncertanities for $ZH$, $W^-H$, and $W^+H$ production is 
shown in~\fig{fig:pdf_uncertainty_comp}.
The uncertainty bands represent the 7-point scale 
variations. For most PDF sets, the scale uncertainty 
decreases with increasing $Q$, while for NNPDF23 it 
decreases up to $Q \sim 1500$ GeV and then rises slowly. 
Differences among PDF sets are below $4\%$ at low $Q$ 
but increase at higher invariant masses. For 
$Q > 2000$ GeV, larger deviations from MMHT2014 are 
seen, especially for ABMP16 and CT18, with the largest 
effect in the $W^-H$ channel, reaching about $20\%$ 
for ABMP16 at high $Q$.
We further evaluate the intrinsic PDF uncertainty using 
the MMHT2014nnlo68cl set by considering all 51 
eigenvector members and computing asymmetric errors 
with the {\tt LHAPDF} prescription. The resulting 
uncertainties, normalized to the central prediction, 
are displayed in~\fig{fig:match_PDFuncertainty}. 
They reach about $5\%$ in the high-$Q$ region.

\begin{figure}[ht!]
	\centerline{
		\includegraphics[width=10.0cm, height=10.0cm]{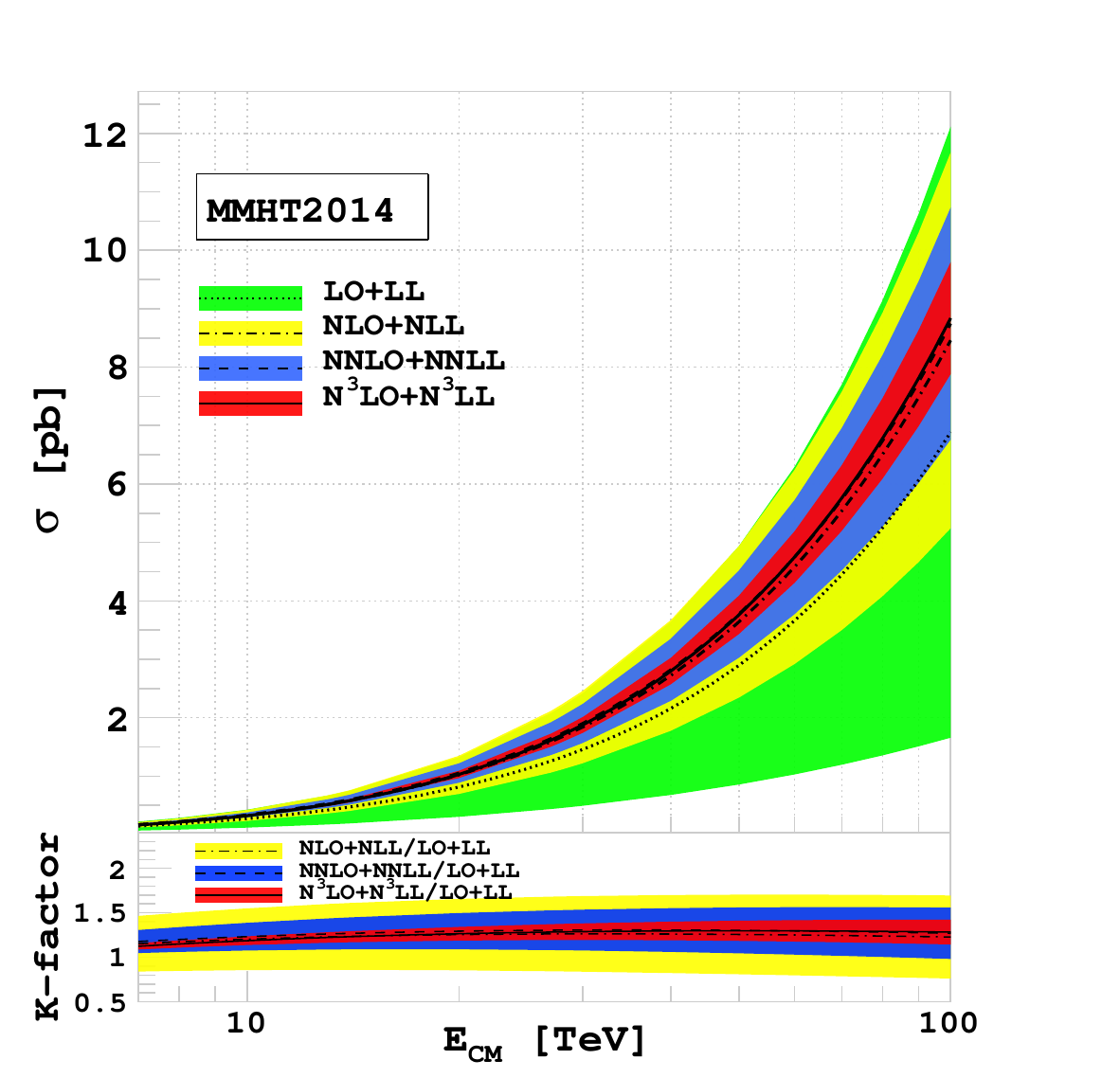}
	}
	\vspace{-2mm}
	\caption{\small{The $b\bar{b} H$ cross section 
        with the corresponding K-factor with respect to LO+LL
	for different center of mass energy at 
	different order.}}
	\label{fig:match_cme_bbH}
\end{figure}

As a final example, we consider Higgs production via 
bottom-quark annihilation at the LHC. This process was 
previously studied up to NNLO+NNLL and later extended to 
N$^3$LO+N$^3$LL as discussed in Ref.~\cite{Das:2022zie}, including 
an analysis of renormalization scale uncertainties. 
Here, we complement those results by performing a full 
7-point scale variation for the $b\bar{b}\to H$ cross section.
The predictions from LO+LL to N$^3$LO+N$^3$LL, together 
with their scale uncertainties, are shown in 
~\fig{fig:match_cme_bbH} for collider energies 
$7 \le \sqrt{S} \le 100$ TeV. At $\sqrt{S}=13.6$ TeV, 
the scale uncertainty decreases from $5.26\%$ at N$^3$LO 
to $4.98\%$ at N$^3$LO+N$^3$LL. In general, resummation 
reduces the scale uncertainty, while at fixed order it 
tends to increase with $\sqrt{S}$.

\section{Summary}\label{sec:conclusion}
We have investigated threshold resummation effects for 
associated Higgs production ($VH$) and Higgs production 
via bottom-quark annihilation ($b\bar{b}H$) up to 
N$^3$LO+N$^3$LL accuracy. Using known resummation 
coefficients and complete N$^3$LO matching from 
\texttt{n3loxs}, we include all regular contributions 
at this order.
For $VH$ production, the resummed corrections beyond 
fixed-order N$^3$LO are small, indicating good 
perturbative convergence. While fixed-order results 
show smaller scale uncertainties at low invariant mass $Q$, 
resummation significantly reduces the 7-point scale 
uncertainties in the high-$Q$ region ($Q \gtrsim 1500$ GeV), 
down to about $0.1\%$.
Overall, scale uncertainties are well under control at 
N$^3$LL accuracy for both $VH$ and $b\bar{b}H$ processes. 
We find that the dominant remaining uncertainty arises from PDFs and
electroweak corrections become important at this 
level of precision.

\end{document}